\documentclass[final,3p,times,twocolumn]{elsarticle}
\usepackage{graphicx}
\usepackage{amssymb}
\usepackage{amsmath}

\usepackage{graphicx}% Include figure files
\usepackage{dcolumn}% Align table columns on decimal point
\usepackage{bm}% bold math
\usepackage{eso-pic}
\usepackage{fix-cm}
\usepackage{natbib}

\newcommand{\rzcm}{cm$^{-1}$}
\newcommand{\rv}[1]{\textcolor[rgb]{0.0,0.00,0.00}{#1}}

\biboptions{comma, sort&compress, square}

\journal{Thin Solid Films}

\begin{document}
\graphicspath{{figures/}}

\begin{frontmatter}

%% Title, authors and addresses

%% use the tnoteref command within \title for footnotes;
%% use the tnotetext command for the associated footnote;
%% use the fnref command within \author or \address for footnotes;
%% use the fntext command for the associated footnote;
%% use the corref command within \author for corresponding author footnotes;
%% use the cortext command for the associated footnote;
%% use the ead command for the email address,
%% and the form \ead[url] for the home page:
%%
%% \title{Title\tnoteref{label1}}
%% \tnotetext[label1]{}
%% \author{Name\corref{cor1}\fnref{label2}}
%% \ead{email address}
%% \ead[url]{home page}
%% \fntext[label2]{}
%% \cortext[cor1]{}
%% \address{Address\fnref{label3}}
%% \fntext[label3]{}

\title{Free-charge carrier parameters of n-type, p-type and compensated InN:Mg determined by Infrared Spectroscopic Ellipsometry}

%% use optional labels to link authors explicitly to addresses:
%% \author[label1,label2]{<author name>}
%% \address[label1]{<address>}
%% \address[label2]{<address>}

\author[UNL]{S. Sch\"{o}che}
\address[UNL]{Department of Electrical Engineering, Center for Nanohybrid Functional Materials, University of Nebraska-Lincoln, U.S.A.}
\ead{schoeche@huskers.unl.edu}
\author[UNL]{T. Hofmann}
%\address{Department of Electrical Engineering, Center for Nanohybrid Functional Materials, University of Nebraska-Lincoln, U.S.A.}
\author[IFM]{V. Darakchieva}
\address[IFM]{Department of Physics, Chemistry and Biology (IFM), Link\"oping University, Sweden}
\author[Chiba]{X. Wang}
\address[Chiba]{Graduate School of electrical and Electronics Engineering and InN-Project as a CREST program of JST, Chiba University, Japan}
\author[Chiba]{A. Yoshikawa}
%\address{Graduate School of electrical and Electronics Engineering and InN-Project as a CREST program of JST, Chiba University, Japan}
\author[Shiga1]{K. Wang}
\author[Shiga1]{T. Araki}
\address[Shiga1]{Research Organization of Science and Engineering, Ritsumeikan University, Shiga, Japan}
\author[Shiga2]{Y. Nanishi}
\address[Shiga2]{Department of Photonics, Ritsumeikan University, Shiga, Japan}
\author[UNL]{M. Schubert}
%\address{Department of Electrical Engineering, Center for Nanohybrid Functional Materials, University of Nebraska-Lincoln, U.S.A.}

\begin{abstract}

Infrared spectroscopic ellipsometry is applied to investigate the free-charge carrier properties of Mg-doped InN films. Two representative sets of In-polar InN grown by molecular beam epitaxy with Mg concentrations ranging from $1.2\times10^{17}$\,cm$^{-3}$ to $8\times10^{20}$\,cm$^{-3}$ are compared. P-type conductivity is indicated for the Mg concentration range of $1\times10^{18}$\,cm$^{-3}$ to $9\times10^{19}$\,cm$^{-3}$ from a systematic investigation of the longitudinal optical phonon plasmon broadening and the mobility parameter in dependence of the Mg concentration. A parameterized model that accounts for the phonon-plasmon coupling is applied to determine the free-charge carrier concentration and mobility parameters in the doped bulk InN layer as well as the GaN template and undoped InN buffer layer for each sample. The free-charge carrier properties in the second sample set are consistent with the results determined in a comprehensive analysis of the first sample set reported earlier [Sch\"oche~\emph{et al.}, J. Appl. Phys. \textbf{113}, 013502 (2013)]. In the second set, two samples with  Mg concentration of $2.3\times10^{20}$\,cm$^{-3}$ are identified as compensated n-type InN with very low electron concentrations which are suitable for further investigation of intrinsic material properties that are typically governed by high electron concentrations even in undoped InN. The compensated n-type InN samples can be clearly distinguished from the p-type conductive material of similar plasma frequencies by strongly reduced phonon plasmon broadening.

\end{abstract}

\begin{keyword}
%% keywords here, in the form: keyword \sep keyword
infrared spectroscopic ellipsometry, Mg-doped InN, p-type, compensation, free-charge carrier parameters
%% MSC codes here, in the form: \MSC code \sep code
%% or \MSC[2008] code \sep code (2000 is the default)

\end{keyword}

\end{frontmatter}

%%
%% Start line numbering here if you want
%%
% \linenumbers

%% main text
\section{Introduction}
\label{Sec:Introduction}

Preparation and control of p-type conductivity is still a challenge towards fabrication of InN-based photonic devices. Several authors have demonstrated that p-type conductivity in InN can be achieved by introduction of Mg as acceptor impurity~\cite{Jones2006, Wang2007, Wang2007a, Wang2007b, Wang2008, Wang2010, Yoshikawa2010}. It is known however that p-type regions in InN are embedded as buried channels in between high-density electron layers that form on the surface and at the interface between InN and substrate (or buffer layer) due to intrinsic defects and elements as oxygen or hydrogen that act as shallow donors~\cite{Gallinat2009,Darakchieva2010}. This behavior affects the confirmation of the conductivity type and impedes the characterization of the p-type region by conventional electrical methods like electrical Hall-effect or hot probe measurements. This difficulty is not only attributed to the high sheet charge density on the surface but also to the large difference between electron (1000-2500\,cm$^2$/Vs) and hole mobility (20-75\,cm$^2$/Vs). Electrolyte capacitance-voltage measurement (ECV) is commonly applied for the determination of hole concentrations while mobility values are estimated by fitting sheet conductivities from Hall-measurements of samples with different thicknesses. Indications for successful p-type doping were only found for Mg concentrations of $1\times10^{18}$\,cm$^{-3}$ to $9\times10^{19}$\,cm$^{-3}$~\cite{Yoshikawa2010,Wang2011}, which is explained by the necessity to overcome a fairly high intrinsic donor concentration caused by lattice defects and residual shallow donors like hydrogen or oxygen~\cite{Gallinat2009,Darakchieva2010} for low Mg concentrations and an increasing defect density for very high Mg concentrations. The influence of the polarity of the buffer layer on growth and physical properties of InN was investigated~\cite{Wang2007a,Wang2007b}. These results were confirmed by other groups~\cite{Wang2011}.

We recently demonstrated the ability of infrared spectroscopic ellipsometry (IRSE) to identify the different carrier types and determine the free-charge carrier parameters concentration and mobility in a series of InN samples (Set~A) with systematically varied Mg concentration~\cite{Schoeche2013}. IRSE measurements were followed by a dedicated model dielectric function (MDF) line shape analysis which provides access to phonon mode and free charge carrier parameters of the group-III nitride thin film sample constituents. In combination with additional NIR-VIS-VUV and optical Hall-effect measurements (generalized ellipsometry with applied external static magnetic field), p-type conductivity in the Mg concentration range $1\times10^{18}$\,cm$^{-3}$$\leq$[Mg]$\leq$$3\times10^{19}$\,cm$^{-3}$ could be concluded from the appearance of a dip structure in the infrared spectral region related to a loss in reflectivity of p-polarized light as a consequence of reduced longitudinal optical (LO) phonon plasmon (LPP) coupling, by vanishing free-charge carrier induced birefringence in the optical Hall-effect measurements, by a sudden change in phonon-plasmon broadening behavior despite continuous change in the Mg concentration, and a strongly reduced mobility for the p-type samples. The free-charge carrier concentration and mobility parameters were determined by applying Kukharskii's model that accounts for the LPP coupling.

The same approach is applied here to investigate the free-charge carrier parameters in a second sample set (Set~B)~\cite{Wang2011} grown in a different reactor \rv{and under a different growth regime (highly In-rich conditions (Set~B) vs. near-stoichiometric regime (Set~A)).} In this sample set, the Mg concentrations were chosen in the range for which p-type conductivity was reported earlier~\cite{Yoshikawa2010} in order to investigate the critical Mg concentrations for transitions between n- and p-type conductivity. In this work, we show that the Mg concentration range for p-type conductivity found in sample Set~B is consistent with the results reported for sample Set~A, i.e., the doping mechanisms are independent on a particular reactor setup \rv{and growth regime}. The determined free-charge carrier parameters for sample Set~B are in excellent agreement with the results reported before for sample Set~A~\cite{Schoeche2013}. In the Set~B, two samples with  Mg concentration of $2.3\times10^{20}$\,cm$^{-3}$ are identified as compensated n-type InN with very low electron concentrations by strongly reduced LPP broadening compared to the p-type samples of similar plasma frequencies. The broadening parameters of the upper and lower LPP branch of the n-type samples in the second sample set follow the expected LPP broadening behavior in dependence of the plasma frequency and prove the expected different broadening behavior for electron and hole phonon-plasmons as pointed out in earlier publication~\cite{Schoeche2013}. The results of sample Set~A as reported in Ref.~\cite{Schoeche2013} are included here for comparison.

\section{Theory}
\label{Sec:Theory}

For the samples discussed here, the optical axes of the materials constituents, which are optically anisotropic with uniaxial optical properties, is oriented parallel to the surface normal. Therefore, without applying a magnetic field during the measurement, no mode conversion of light polarized parallel \,(p) to polarization perpendicular\,(s) to the plane of incidence and vice versa will occur and standard ellipsometry can be applied~\cite{Schubert04}. The standard ellipsometric parameters $\Psi$ and $\Delta$ are defined by the ratio $\rho$ of the complex valued Fresnel reflection coefficients
\begin{equation}
	\rho=\frac{r_p}{r_s}=\tan{\Psi}\cdot\exp(\mathrm{i}\Delta).
\end{equation}

The ellipsometry data was analyzed by using a layer-stack model including the sapphire substrate, the GaN template layer~\cite{SchubertIRSEBook_2004}, the undoped InN buffer layer, and the Mg doped bulk InN layer. The light propagation within the entire sample stack is calculated by applying a $4\times4$ matrix algorithm for multilayer systems assuming plane parallel interfaces~\cite{SchubertIRSEBook_2004}. In order to reduce parameter correlation, the measurements are obtained at multiple angles of incidence and analyzed simultaneously. A regression analysis\,(Levenberg-Marquardt algorithm) is performed, where the model parameters are varied until calculated and experimental data match as close as possible~\cite{Fujiwara_book2007}. The sapphire model dielectric function (MDF), uncoupled phonon mode parameters, and $\varepsilon_{\infty}$ values for GaN and InN were taken from our previous work~\cite{Darakchieva2009,SchubertIRSEBook_2004,Schubert2000,Kasic2000}. The contribution of free-charge carriers to the MDF and phonon-plasmon coupling was described by applying Kukharskii's model~\cite{Kukharskii73,SchubertIRSEBook_2004,Kasic2000}:
\begin{align}
\varepsilon_{\parallel,\perp}(\omega)=\varepsilon_{\parallel,\perp;\infty}&\frac{(\omega^2+\mathrm{i}\gamma_{\mathrm{LPP}^{-};\parallel,\perp}\omega-\omega^2_{\mathrm{LPP}^{-};\parallel,\perp})}{\omega(\omega+\mathrm{i}\gamma_{\mathrm{p};\parallel,\perp})} \nonumber\\
\cdot&\frac{(\omega^2+\mathrm{i}\gamma_{\mathrm{LPP}^{+};\parallel,\perp}\omega-\omega^2_{\mathrm{LPP}^{+};\parallel,\perp})}{(\omega^2+\mathrm{i}\gamma_{\mathrm{TO};\parallel,\perp}\omega-\omega^2_{\mathrm{TO};\parallel,\perp})},
\end{align}
where $\omega_{\mathrm{LPP}^{-/+}}$ are the frequencies of the LPP$^-$/LPP$^+$ branches and $\gamma_{\mathrm{LPP}^{-/+}}$ the according LPP broadening parameters. $\gamma_{\mathrm{p}}$ is the plasma broadening parameter which is connected to the effective mass m$^{\ast}$ and the optical mobility $\mu$ by $\gamma_{\mathrm{p}}=q/m^{\ast}\mu$, with $\left|q\right|=e$ being the elementary charge. $\omega_{\mathrm{LPP}^{-/+}}$ can be extracted from
\begin{equation}
\omega_{\mathrm{LPP}^{-/+}}=\sqrt{\frac{1}{2}\left[\omega^2_{\mathrm{p}}+\omega^2_{\mathrm{LO}}\pm\sqrt{(\omega^2_{\mathrm{p}}+\omega^2_{\mathrm{LO}})^2-4\omega^2_{\mathrm{p}}\omega^2_{\mathrm{TO}}}\right]},
\end{equation}
with $\omega_{\mathrm{TO/LO}}$ being the uncoupled transverse optical (TO)/LO phonon mode frequencies and $\omega_{\mathrm{p}}$ the plasma frequency given by $\omega_{\mathrm{p}}^2=Nq^2/m^{\ast}\varepsilon_0\varepsilon_{\infty}$. \textit{N} is the free-charge carrier concentration. The plasma frequency $\omega_{\mathrm{p}}$ and plasma broadening parameter $\gamma_{\mathrm{p}}$ provide access only to the coupled quantities $N/m^{\ast}$ and $\mu{m^{\ast}}$. Conclusions about carrier concentration and mobility from the IRSE measurements are therefore only possible by assuming a value for the effective mass of electrons or holes, respectively. Further, the sign of the charge $q$ is not accessible without applying an additional magnetic field during the measurement, i.e. direct determination of the carrier type is not possible from IRSE. However, the broadening parameters $\gamma_{\mathrm{LPP}^{-/+}}$ are independent of the effective mass of the carriers and directly obtained from the model analysis. As discussed in previous publication, holes are expected to reveal much lower mobility and thus higher broadening parameters $\gamma_{\mathrm{LPP}^{-/+}}$ compared to electrons~\cite{Schoeche2013}. Thus, a different broadening behavior is expected for p-type conducting samples compared to n-type conducting samples and the LPP broadening parameter can be taken as indicator for a change of the conductivity type~\cite{Schoeche2013}. In order to achieve the best match between experimental and model data free-charge carrier contributions in the doped bulk InN layer, the GaN template layer and the not intentionally doped InN buffer layer were included in the model analysis.

\rv{The band structure of InN could so far only be studied theoretically due to the lack of intrinsic InN samples~\cite{Rinke2008}. All unintentionally doped InN samples reported in literature show n-type conductivity. Consequently, Burstein-Moss effect calculations always need to be applied in order to extract the intrinsic band-gap parameters. The conduction band of InN was found to be non-parabolic, i.e. the effective electron mass parameter depends on the free-electron concentration~\cite{Hofmann2008}. An extrapolated value for the effective electron mass parameter for intrinsic InN of $m^\ast=0.04\,m_0$ with negligible anisotropy at the $\Gamma$-point was reported~\cite{Hofmann2008}. The splitting between the $\Gamma_{9v}$ heavy-hole valence band and the $\Gamma_{7v}^{\mathrm{so}}$ spin-orbit split-off band is too small too be experimentally resolved~\cite{Fujiwara2008}. For the crystal-field splitting a small positive value of $\Delta_{\mathrm{CR}}=0.065\,$eV is predicted theoretically~\cite{Rinke2008}. Fujiwara~\emph{et al.} estimated that $80\%$ of the free holes will occupy heavy hole states~\cite{Fujiwara2008}. Therefore, mainly heavy holes are probed by our ellipsometry experiment. An estimation of heavy and light hole mass parameters from acceptor activation energies determined by IRR, IR transmission and photoluminescence at low temperature was reported recently~\cite{Fujiwara2011}. Contributions of electronic interband transitions between the valence bands were not reported for spectroscopic ellipsometry measurements on group-III nitride semiconductors and would be expected to be outside the accessible spectral range due to the small valence band splittings in InN.}

\section{Experiment}
\label{Sec:Experiment}

Two sets of InN films with varying Mg concentrations grown on c-plane GaN/sapphire templates by molecular beam epitaxy were investigated. \rv{Extended defects and impurity incorporation in InN films are expected to strongly depend on growth temperature and III/V ratio~\cite{Yoshikawa2010}. Even samples grown in one reactor under different conditions can show significantly different film properties~\cite{Yoshikawa2010}. For Set~A near-stoichiometric growth conditions at slightly higher III/V ratio were chosen in order to avoid the formation of In droplets~\cite{Yoshikawa2010}. For Set~B the so-called droplet elimination by nitrogen radical irradiation (DERI) method was applied which utilizes the generally higher crystalline quality in InN films grown at strongly In-rich conditions and uses nitrogen radical irradiation to eliminate the In droplets formed under this regime~\cite{Yamaguchi2009}. Details on the reactor designs and exact growth conditions can be found elsewhere~\cite{Yoshikawa2010,Yamaguchi2009}. Since the growths regimes differ significantly from each other, similar sample properties upon introduction of doping atoms are not obvious.}

Each set included an undoped InN reference sample and Mg-doped InN with systematically increased Mg concentration ranging from $1.2\times10^{17}$\,cm$^{-3}$ to $8.0\times10^{20}$\,cm$^{-3}$ (Set~A) and from $1.0\times10^{18}$\,cm$^{-3}$ to $2.3\times10^{20}$\,cm$^{-3}$ (Set~B) in order to achieve p-type doping. For all samples, the InN growth started with a 50\,nm undoped layer on top of which about 400 nm thick layers of doped InN were deposited. P-type conductivity was proven by ECV measurements in Set~A for Mg concentrations between $1.1\times10^{18}$\,cm$^{-3}$ and $2.9\times10^{19}$\,cm$^{-3}$ by Yoshikawa~\emph{et al.}~\cite{Yoshikawa2010} and in Set~B for Mg concentrations between $9.0\times10^{18}$\,cm$^{-3}$ and $9.0\times10^{19}$\,cm$^{-3}$ by Wang~\emph{et al.}~\cite{Wang2011}. A summary of the samples in each set is given in Tab.~\ref{Tab:samples} with the conductivity type indicated in parenthesis. For sample Set~B, two different growth conditions (N-rich and In-rich) were applied during the deposition for the Mg concentration at which the transition from p-type to n-type was expected.

The IRSE measurements have been carried out using a commercial Fourier transform-based IR ellipsometer\,(J.A.~Woollam~Co.,~Inc.) in the spectral range from 300\,cm$^{-1}$ to 6000\,\rzcm. The details on the IRSE data analysis of sample Set~A were described in Ref.~\cite{Schoeche2013} and the results are included here for comparison.

\begin{table}[tbh]
\caption{Summary of the two sample sets. The conductivity type as determined from IRSE is indicated in parenthesis. Note, that in Set~B two samples have the same Mg concentration but are grown under slightly different conditions. Set A was investigated in Ref.~\cite{Schoeche2013} and results are included here for comparison.}
	\centering
		\begin{tabular}{cccc}
		\hline\\[-2.3ex]
			\multicolumn{2}{c}{\textbf{Set~A}} &\multicolumn{2}{c}{\textbf{Set~B}}\\
			Sample& [Mg] (cm$^{-3})$& Sample& [Mg] (cm$^{-3})$\\
		\hline
		\hline \\[-2.3ex]
		A1&undoped (n)&B1&undoped (n)\\
		A2&$1.2\times10^{17}$ (n)&B2&$1.0\times10^{18}$ (n)\\
		A3&$1.0\times10^{18}$ (p)&B3&$9.0\times10^{18}$ (p)\\
		A4&$5.5\times10^{18}$ (p)&B4&$2.2\times10^{19}$ (p)\\
		A5&$2.9\times10^{19}$ (p)&B5&$5.6\times10^{19}$ (p)\\
		A6&$1.8\times10^{20}$ (n)&B6&$9.0\times10^{19}$ (p)\\
		A7&$8.0\times10^{20}$ (n)&B7&$2.3\times10^{20}$ (n)\\
        & & & (N-rich)\\
        &  &B8&$2.3\times10^{20}$ (n)\\
        & & & (In-rich)\\
		\hline
		\end{tabular}
	\label{Tab:samples}
\end{table}

\section{Results and Discussion}
\label{Sec:Results}

\begin{figure}[tb]
\centerline{\includegraphics[keepaspectratio=true,trim=2 0 0 0, clip, width=0.49\textwidth]{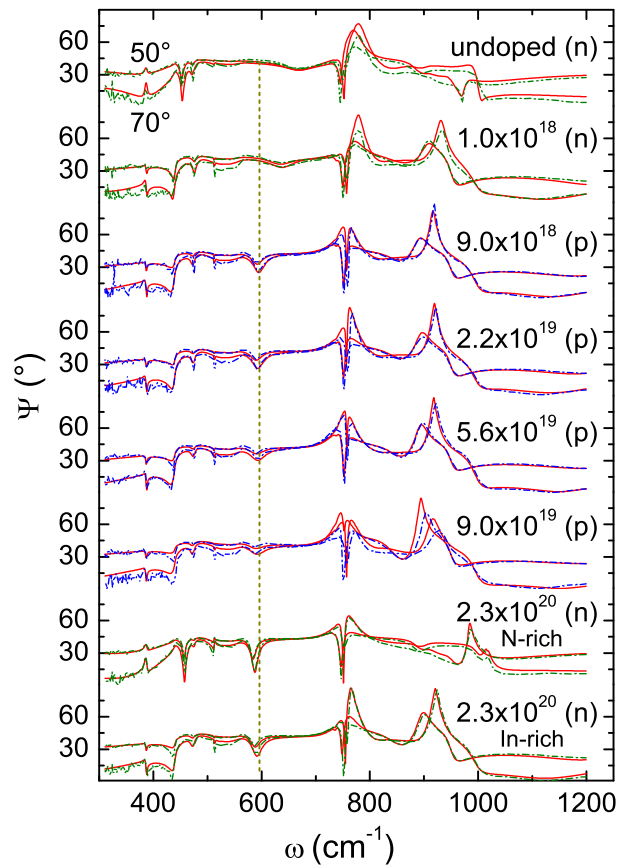}}
\caption{Comparison of experimental (dashed lines) and best-matching model data (solid lines) of IRSE measurements for the samples of Set~B. The Mg concentration in each sample is given and conductivity type as determined from IRSE is indicated in parenthesis. The characteristic dip around 600\,\rzcm due to the reduced LPP shift and broadening of the InN A$_1$-LPP mode for the p-type conducting samples is marked by the vertical line. \rv{Note, the significant change in the line shape of the undoped sample and [Mg]$=2.3\times10^{20}$\,cm$^{-3}$ (N-rich) around 400\,cm$^{-1}$ and above 1000\,cm$^{-1}$ is related to a significantly different thickness of the GaN template layer (2\,$\mu$m vs. 3.5\,$\mu$m).}}
\label{Fig:IRSE}
\end{figure}

Experimental and best-matching model data for the samples of Set~B are given in Fig.~\ref{Fig:IRSE}. A detailed description of the different features in the spectrum can be found in Ref.~\cite{Schoeche2013}. The spectra presented here are similar to the ones of sample Set~A investigated in the same reference showing the characteristic dip structure around 600\,cm$^{-1}$ related to a reduced LPP shift for the p-type samples. The spectral position of this dip is related to the plasma frequency, i.e., the free-charge carrier concentration in the sample, while the shape is mainly determined by the LPP broadening parameters and the mobility~\cite{Schoeche2013}. The most significant difference compared to the data of sample Set~A is the occurrence of a deep trench at about 600\,cm$^{-1}$in the two samples with a Mg concentration of $2.3\times10^{20}$\,cm$^{-3}$. The analysis of the broadening behavior and mobility parameters suggests that these two samples are n-type conductive with a strongly reduced number of free-charge carriers, i.e., the samples are highly compensated n-type InN.

\begin{figure*}[tb]
\centerline{\includegraphics[keepaspectratio=true,trim=0 0 0 0, clip, width=0.47\textwidth]{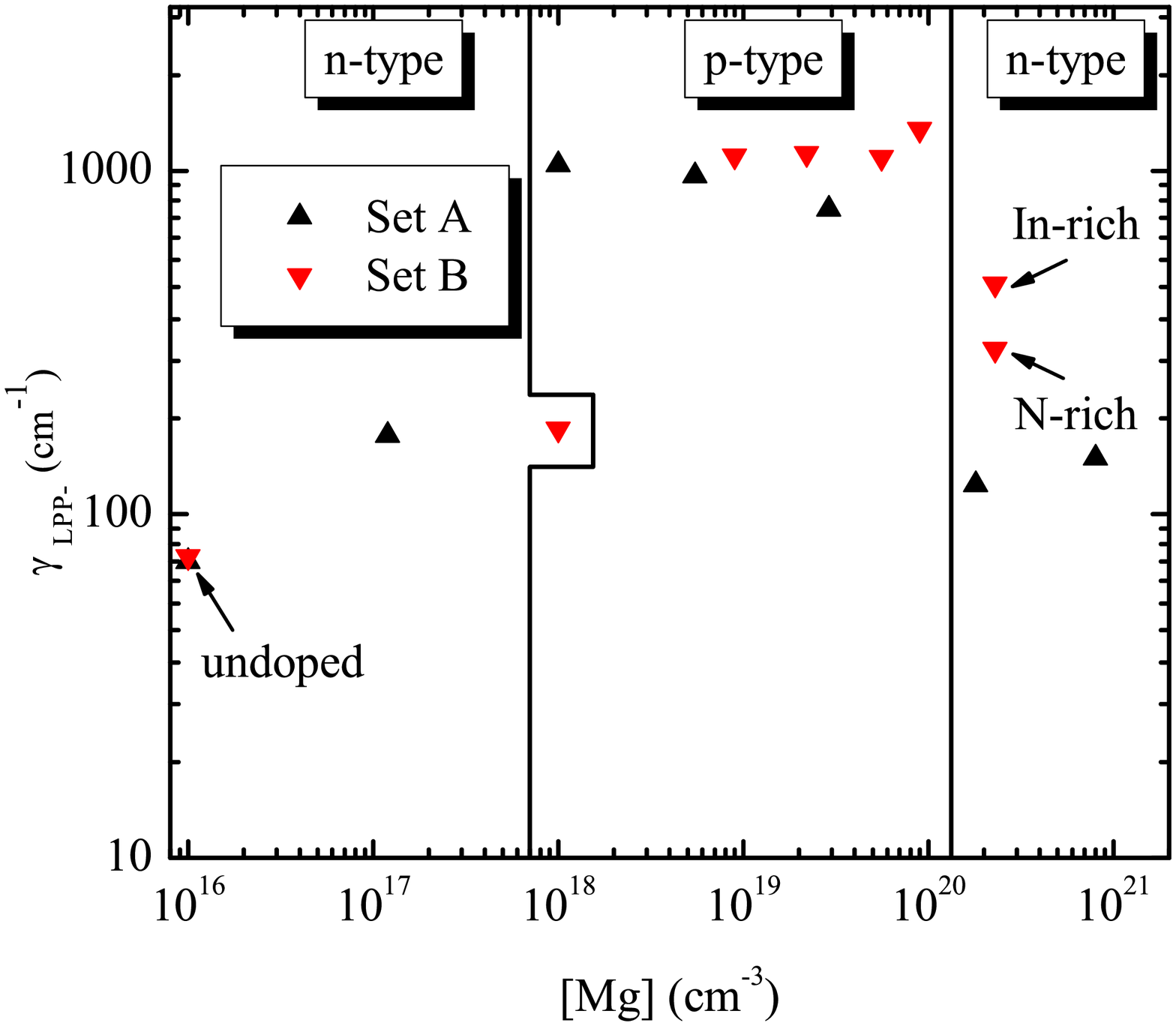}\hfill  \includegraphics[keepaspectratio=true,trim=0 0 0 0, clip, width=0.47\textwidth]{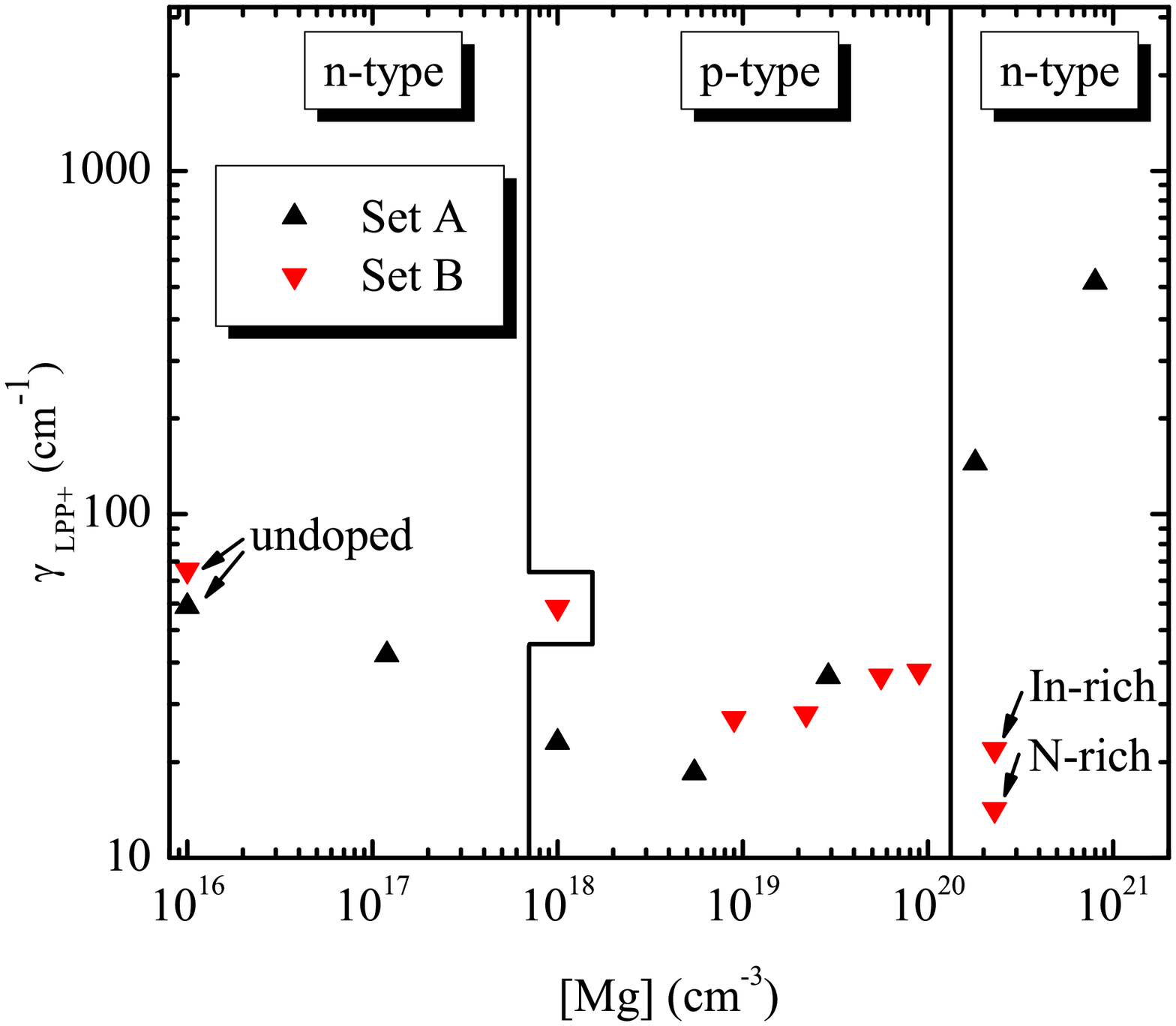}}
\caption{Comparison of the best-match model parameters for the LPP broadening parameters for the lower LPP$^-$ (left) and upper LPP$^+$ (right) branch as obtained from the regression analysis for both sample sets. Values for the undoped reference samples are indicated on the coordinate axis on the left at [Mg]$=1\times10^{16}$\,cm$^{-3}$.}
\label{Fig:gamma}
\end{figure*}

\begin{figure}[bh!]
\centerline{\includegraphics[keepaspectratio=true,trim=0 0 0 0, clip, width=0.45\textwidth]{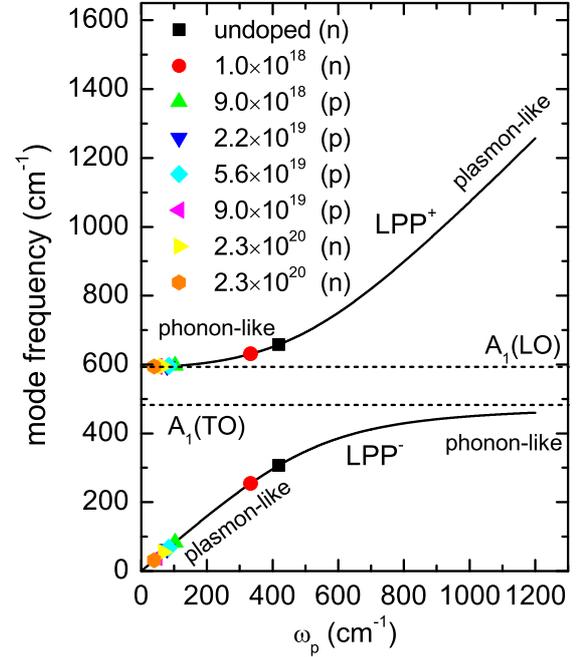}}
\caption{LPP branches in dependence of the plasma frequency $\omega_{\mathrm,{p}}$. The position of the LPP$^{+/-}$ modes is indicated by symbols for each sample of Set~B as obtained from the model analysis.}
\label{Fig:LPP}
\end{figure}

The LPP broadening parameters for the upper (LPP$^+$) and lower (LPP$^-$) phonon plasmon branch determined from the best-match model analysis of both sample sets are compared in Fig.~\ref{Fig:gamma}. The transition from n-type to p-type conductivity occurs at slightly higher Mg concentrations in sample Set~B compared to Set~A. The broadening parameters of the p-type samples in Set~B match the values found for Set~A and the same sudden change of broadening behavior despite continuous increase of the Mg concentration as reported for Set~A is observed allowing to identify the transitions between the conductivity types. The two samples identified as compensated material show completely different broadening behavior compared to all other samples in both sets. The broadening parameter of the LPP$^+$ branch for the compensated samples is much lower than for all n-type samples in both sets and even lower than the broadening parameter of the p-type samples in the same set. The broadening parameter of the LPP$^-$ branch for the two compensated samples is slightly higher than for the n-type samples of comparable Mg concentration in Set~A, but much lower than the broadening parameters of the p-type samples.
This behavior can be understood by taking into account the LPP dispersion in dependence of the plasma frequency (Fig.~\ref{Fig:LPP}). The frequencies of the LPP branches for each sample of Set~B according to the determined plasma frequencies are marked on the dispersion curves. For the upper branch, the broadening behavior is expected to change from plasmon-like (large broadening) to phonon-like (small broadening) with decreasing plasma frequency. The opposite behavior would be expected for the lower branch. The comparison of the compensated samples with the other n-type samples of Set~B shows exactly this behavior. The largest plasma frequency was determined for the undoped samples, a slightly smaller plasma frequency is found for the n-type sample of lowest Mg concentration, and the lowest plasma frequency was determined for the compensated samples. Accordingly, the LPP$^+$ broadening parameter systematically decreases while the LPP$^+$ parameter decreases. This trend could not be observed in sample Set~A because of the absence of n-type samples of very low free-charge carrier concentrations. Note, that even though similar plasma frequencies as for the compensated samples are also determined for the p-type samples, much smaller broadening values for LPP$^+$ as well as LPP$^-$ are determined for the compensated n-type samples. This is consistent with the expected smaller phonon-plasmon broadening for electrons as compared to holes due to the strongly reduced hole mobility.

\begin{figure*}[tb]
\centerline{\includegraphics[keepaspectratio=true,trim=0 0 0 0, clip, width=0.47\textwidth]{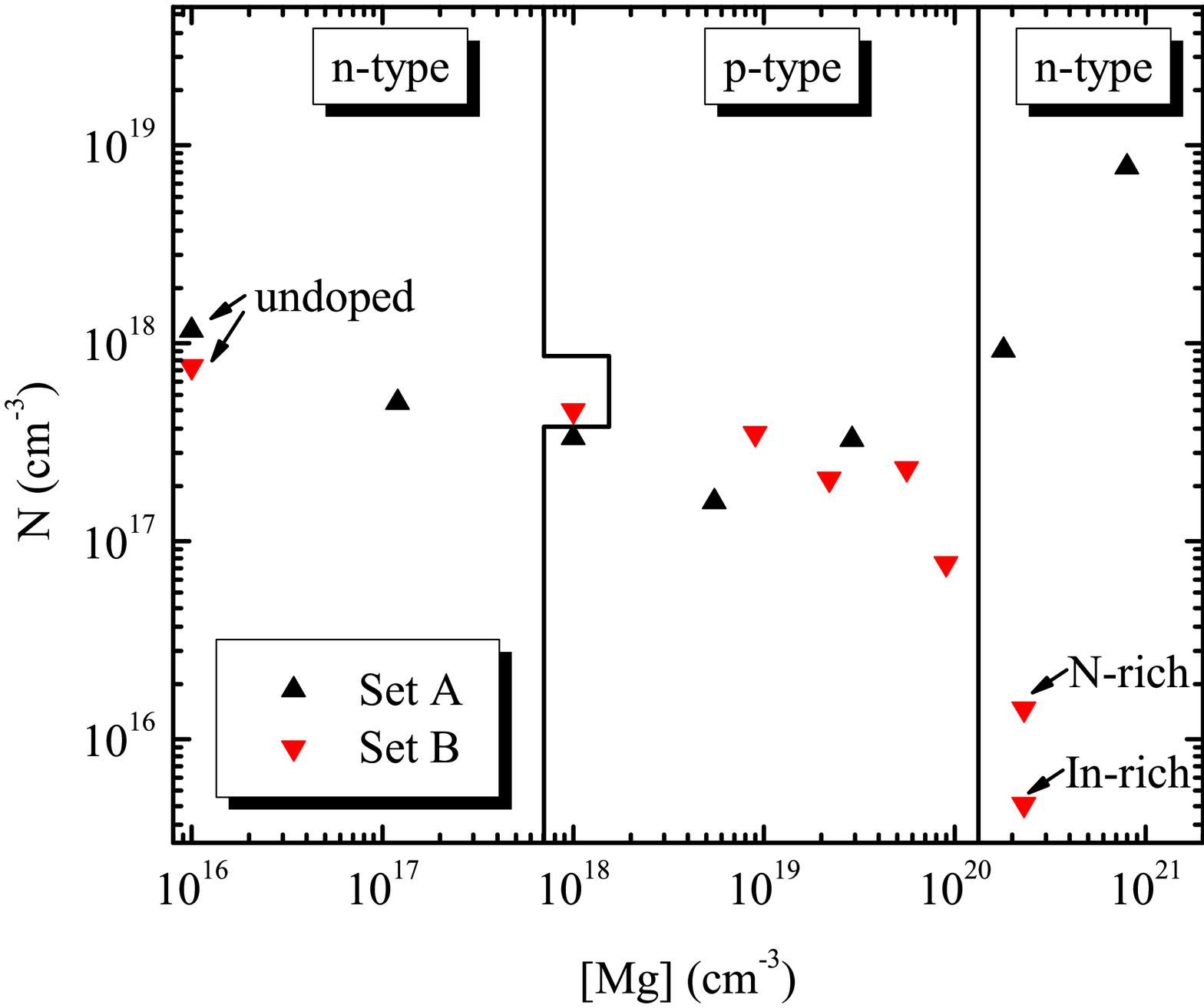} \hfill \includegraphics[keepaspectratio=true,trim=0 0 0 0, clip, width=0.47\textwidth]{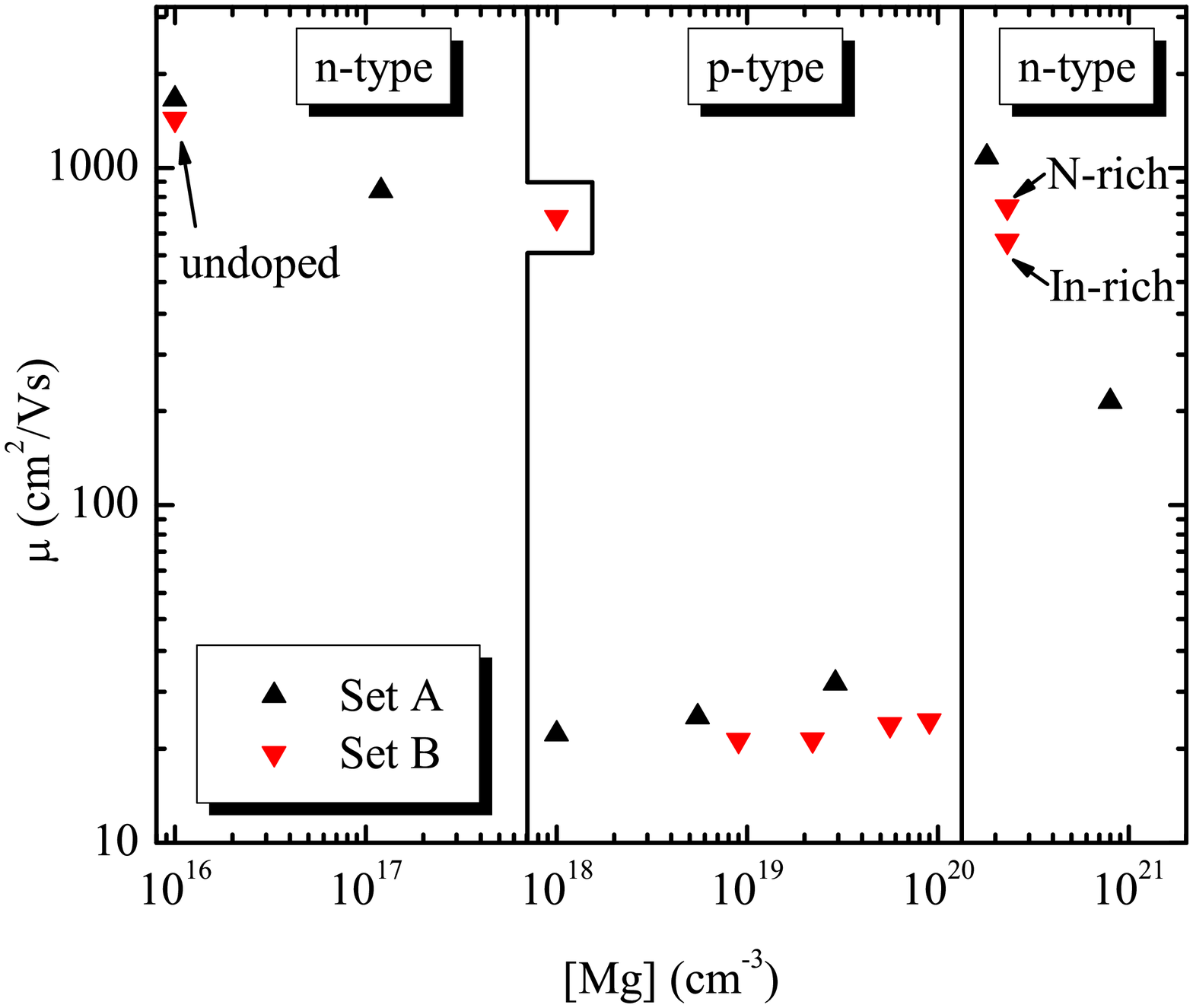}}
\caption{Comparison of the best-match model parameters for the free-charge carrier concentration $N$ (left) and mobility $\mu$ (right) as determined from the IRSE model analysis for both sample sets. Values for the undoped reference samples are indicated at the left at [Mg]$=1\times10^{16}$\,cm$^{-3}$.}
\label{Fig:N_Mu}
\end{figure*}

Assuming an effective hole mass of 0.42\,$m_0$ for the p-type samples~\cite{Fujiwara2008} and using the dependence of the electron effective mass on the carrier concentration as determined by OHE~\cite{Hofmann2008}, it is therefore now possible to deduce the parameters of the free-charge carrier concentration $N$ and the mobility $\mu$ from $\omega_{\mathrm{p}}$ and $\gamma_{\mathrm{p}}$. The comparison of the derived parameters for both sample sets is shown in Fig.~\ref{Fig:N_Mu}. The free-charge carrier concentrations in Set~B are comparable to the values found for Set~A with the already discussed exception of the two compensated samples. The mobility values determined for sample Set~B are in excellent agreement with the values of Set~A. The very high mobility values determined for the samples with a Mg concentration of $2.3\times10^{20}$\,cm$^{-3}$ further support the identification as compensated n-type material.   A higher mobility value is determined for the compensated sample grown under slightly N-rich conditions compared to the one grown under slightly In-rich conditions. This result is in agreement with the lower broadening parameters for upper as well as lower LPP branch which might indicate a better crystalline quality of the sample grown under N-rich conditions. ECV measurements independently confirmed the n-type character of these samples~\cite{Wang2011}. Note, that the sample B6 with a Mg concentration of $9.5\times10^{19}$\,cm$^{-3}$ was identified as n-type material by ECV and thermopower measurements in Ref.~\cite{Wang2011}. However, the interpretation of the ellipsometry analysis suggests p-type behavior of the volume free-charge carriers in this sample. A reduced number of holes compared to the other p-type samples in this set is determined which could result in an increased influence of the n-type buffer layer on the electrical measurements and might explain the ECV and thermopower results.

\section{Conclusion}
\label{Sec:Conclusion}

The free-charge carrier parameters of two sets of Mg-doped InN grown independently by two different groups by MBE \rv{under two different growth regimes} were determined by applying infrared spectroscopic ellipsometry. \rv{While the samples of Set~A were grown under nearly stoichiometric conditions, nitrogen radical irradiation is utilized to eliminate the In droplets formed under highly In-rich growth conditions.} In both sets, the Mg concentration was systematically increased in order to produce p-type conductive material. The ellipsometry data was analyzed by applying Kukharskii's model which accounts for the LO phonon plasmon coupling in material with high free-charge carrier concentration. The transition from n-type to p-type conductivity can be identified in the IRSE analysis by a sudden strong change in the broadening behavior despite a continuous change in the Mg concentration and huge change of the mobility values. The validity of this interpretation was demonstrated in an earlier publication by applying additional NIR-VIS-VUV spectroscopic ellipsometry measurements and optical Hall-effect measurements on sample Set~A~\cite{Schoeche2013} and is in agreement with independent ECV and thermopower measurements~\cite{Yoshikawa2010,Wang2011}.

The transition from n-type to p-type conductivity in sample Set~B is found to occur at slightly higher Mg concentrations than in Set~A, but overall the determined Mg concentration range for p-type doping for sample Set~B is consistent with the results reported earlier for sample Set~A~\cite{Schoeche2013}. This result demonstrates the independence of the doping mechanisms on a particular reactor setup \rv{or growth regime}. The determined parameters for LPP broadening, hole concentration, and mobility for the p-type samples in both sample sets are in excellent agreement. In Set~B, two samples are identified as compensated n-type InN. Samples of comparable low free-charge carrier concentration were so far not available since high intrinsic electron concentrations are typically even found in undoped InN. Doping with suitable amounts of Mg might therefore be a promising method to produce not only p-type material but also compensated material suitable for investigation of intrinsic properties of InN typically governed by high free-charge carrier concentrations. Even though similar values of the plasma frequency are determined for the p-type and compensated material, much lower broadening of upper and lower phonon plasmon branch are determined for the compensated samples which proves the expectation that electron phonon plasmons possess smaller broadening than hole phonon plasmons due to the strongly reduced mobility of holes compared to electrons. The determined mobility values for the compensated samples match the values for n-type samples of higher electron concentration which supports our interpretation.

\section{Acknowledgment}

The authors acknowledge financial support from the National Science Foundation under awards MRSEC DMR-0820521, MRI DMR-0922937, DMR-0907475, and EPS-1004094, by the Swedish Research Council (VR) under grant No.2010-3848, the Swedish Governmental Agency for Innovation Systems (VINNOVA) under the VINNMER International Qualification program, grant No. 2011-03486 and by FCT Portugal under contract PTDC/FIS/100448/2008, and program Ci\^{e}ncia 2007.

%% The Appendices part is started with the command \appendix;
%% appendix sections are then done as normal sections
%% \appendix

%% \section{}
%% \label{}

%% References
%%
%% Following citation commands can be used in the body text:
%% Usage of \cite is as follows:
%%   \cite{key}         ==>>  [#]
%%   \cite[chap. 2]{key} ==>> [#, chap. 2]
%%

%% References with bibTeX database:

\bibliographystyle{model1a-num-names}
%\bibliography{C:/Users/blank/Literatur/BIBTEX/CompleteLibrary}

%% Authors are advised to submit their bibtex database files. They are
%% requested to list a bibtex style file in the manuscript if they do
%% not want to use elsarticle-num.bst.

%% References without bibTeX database:

% \begin{thebibliography}{00}

%% \bibitem must have the following form:
%%   \bibitem{key}...
%%

% \bibitem{}

% \end{thebibliography}

\end{document}